%% file: semi_blind_arxiv.tex
\documentclass{article}
\usepackage{spconf,amsmath,graphicx}

\usepackage{balance}
\usepackage{cite}
\usepackage{amsmath,amssymb,amsfonts}
\input{def}

\usepackage[acronym,shortcuts]{glossaries}
\newacronym{GMM}{GMM}{Gaussian mixture model}
\newacronym{MSE}{MSE}{mean-square-error}
\newacronym{CSI}{CSI}{channel state information}
\newacronym{CME}{CME}{conditional mean estimator}
\newacronym{ML}{ML}{maximum likelihood}
\newacronym{SNR}{SNR}{signal-to-noise ratio}
\newacronym{BS}{BS}{base station}
\newacronym{PDF}{PDF}{probability density function}
\newacronym{LMMSE}{LMMSE}{linear minimum mean square error}

\title{Data-aided Channel Estimation utilizing Gaussian Mixture Models
}

\name{Franz Weißer, Nurettin Turan, Dominik Semmler, and Wolfgang Utschick
\thanks{This work was supported by the Federal Ministry of Education and Research of Germany in the programme of “Souverän. Digital. Vernetzt.”. Joint project 6G-life, project identification number: 16KISK002}
}
\address{TUM School of Computation, Information and Technology, Technical University of Munich, Germany\\
Email: \{franz.weisser, nurettin.turan, dominik.semmler, utschick\}@tum.de}
\usepackage{fancyhdr}

\fancypagestyle{cfooter}{ %
	\fancyhf{} 
	\cfoot{
		This work has been submitted to the IEEE for possible publication. Copyright may be transferred without notice, after which this version may no longer be accessible.}
	
}

\begin{document}
\ninept
\maketitle
\thispagestyle{cfooter}
\begin{abstract}
	
	In this work, we propose two methods that utilize data symbols in addition to pilot symbols for improved channel estimation quality in a multi-user system, so-called semi-blind channel estimation. To this end, a subspace is estimated based on all received symbols and utilized to improve the estimation quality of a Gaussian mixture model-based channel estimator which solely uses pilot symbols for channel estimation. Both of the proposed approaches allow for parallelization. Even the precomputation of estimation filters, which is beneficial in terms of computational complexity is enabled by one of the proposed methods. 
	Numerical simulations for real channel measurement data available to us, show that the proposed methods outperform the studied state-of-the-art channel estimators.
\end{abstract}
%

\input{intro}

\input{system}
\input{method}
\input{estimators}

\input{simulations}
\input{conclusion}

\newpage
\balance
\bibliographystyle{IEEEbib}
\bibliography{mybib}

\end{document}

%% file: def.tex
\usepackage{array}
\usepackage[caption=false,font=footnotesize]{subfig}
\usepackage{stfloats}
\usepackage{url}
\usepackage{graphicx}
\usepackage{textcomp}
\usepackage{xcolor}
\usepackage{tikz}
\usepackage{ellipsis}
\usetikzlibrary{calc}
\usetikzlibrary{decorations.pathreplacing,decorations.markings,shapes.geometric}
\usetikzlibrary{calc,patterns,angles,quotes,shapes,arrows.meta}
\usetikzlibrary{chains}

\usepackage[utf8]{inputenc}

\usepackage{mathtools}
\usepackage{bm}
\usepackage{etoolbox}
\usepackage{scalerel}

\usepackage{fancyhdr}

\newcommand{\vh}{{\bm{h}}}

\newcommand{\vn}{{\bm{n}}}

\newcommand{\vv}{{\bm{v}}}

\newcommand{\vx}{{\bm{x}}}
\newcommand{\vy}{{\bm{y}}}

\newcommand{\mc}{{\bm{C}}}
\newcommand{\md}{{\bm{D}}}

\newcommand{\mh}{{\bm{H}}}

\newcommand{\mn}{{\bm{N}}}

\newcommand{\matp}{{\bm{P}}}

\newcommand{\mv}{{\bm{V}}}

\newcommand{\mx}{{\bm{X}}}
\newcommand{\my}{{\bm{Y}}}

\newcommand{\vmu}{\bm{\mu}}

\newcommand{\tr}{\mathrm{tr}}

\newcommand{\He}{\mathrm{H}}
\newcommand{\T}{\mathrm{T}}

\newcommand{\I}{\mathbf{I}}

\newcommand{\NC}{\mathcal{N}_\mathbb{C}}

\tikzstyle{block} = [draw, rectangle, 
minimum height=4em, minimum width=4em]
\tikzstyle{input} = [coordinate]
\tikzstyle{output} = [coordinate]
\tikzstyle{pinstyle} = [pin edge={to-,thin,black}]

\usetikzlibrary{positioning}

\tikzset{radiation/.style={{decorate,decoration={expanding waves,angle=90,segment length=5pt}}}}

\usetikzlibrary{spy}
\usepackage{pgfplots}
\usepackage{wrapfig}
\usetikzlibrary{arrows,shapes}
\usetikzlibrary{positioning,shapes.callouts}
\usepgfplotslibrary{groupplots,dateplot}
\usetikzlibrary{patterns,shapes.arrows}
\pgfplotsset{compat=newest}

\def\BibTeX{{\rm B\kern-.05em{\sc i\kern-.025em b}\kern-.08em
		T\kern-.1667em\lower.7ex\hbox{E}\kern-.125emX}}

%% file: intro.tex
\section{Introduction}

In modern wireless communication systems, accurate channel estimation is crucial for achieving high data rates and robust transmissions~\cite{Rusek2013,Kabalci2019}. 
The communication link between transmitter and receiver is characterized by its time-varying and frequency-selective nature.
Impairments introduced through multipath propagation, fading, and noise directly impact the quality of the channel. 
Consequently, accurately estimating the \ac{CSI} is pivotal for successfully transmitting data.



Pilot-based channel estimation techniques rank among the most widely adopted methods in wireless communication. These methods involve transmitting pilot signals, known beforehand at the transmitter and receiver side, across the channel using some of the radio resource blocks. Subsequently, the receiver leverages the observed signals to compute a reliable \ac{CSI} estimate.
Unlike pilot-based methods, data-aided techniques, classically referred to as semi-blind, capitalize on the information embedded within the observed data symbols at the receiver to infer channel characteristics. 
These methods exploit structure and redundancy within the transmitted data and yield more robust \ac{CSI} estimates.
The benefit was first studied in~\cite{DeCarvalho1997a} and later adapted for specific tasks in~\cite{Liu2012, Ma2014}. A different view has recently been reconsidered and re-proposed in~\cite{Khan2023}, where reliable decoded data symbols were interpreted as additional pilots.

Over the past few years, machine learning methods have been introduced to improve various tasks in wireless communication~\cite{Ye2018, Nachmani2018, Soltani2019,Shlezinger2021,Weisser2023}, where~\cite{Shlezinger2021,Weisser2023} are examples of so-called model-based machine learning methods.
The idea of machine learning is to enhance the task at hand by using prior information obtained during the learning stage.
For a given \ac{BS} cell, the \ac{PDF} representing potential user channels can be considered valuable prior information.
Since this true underlying distribution is unknown, machine learning methods rely on a representative data set, which is assumed to be available at the \ac{BS}.
Recently, powerful examples of leveraging such prior information were presented in~\cite{Guo2022, Koller2022, Turan2023}.
One approach involves the construction of a \ac{GMM} given a training dataset, in order to capture the \ac{PDF} of the BS cell.
The learned \ac{GMM} enables channel estimation in \cite{Koller2022} or a limited feedback scheme in \cite{Turan2023}.

\emph{Contributions:} In this work, the local Gaussian approximation of the actual \ac{PDF} via a \ac{GMM} and the corresponding formulation of input dependent conditional \ac{LMMSE} estimators from \cite{Koller2022} is adapted to support data-aided channel estimation. To this end, we first depict how the \ac{GMM}-based channel estimator can be utilized to solve a subspace estimation problem. As an alternative, we present a projection method that is computationally more efficient since it allows for the pre-calculation of \ac{LMMSE} filters.
Extensive simulations show the superior performance of our proposed adaptions as compared to state-of-the-art channel estimation approaches in typical massive multiple-input multiple-output (MIMO) systems with multiple users.

%% file: system.tex
\section{System and Channel Model}

We consider a multi-user uplink system with $J$ single-antenna users and a \ac{BS} equipped with $M$ receive antennas. 
The received signal vector at time instance $n$ can then be expressed as
\begin{align}
	\vy(n) 
	&=  \mh \vx(n) + \vn(n) , \quad n=1,...,N
\end{align}
where $\vx(n) = [x_1(n),...,x_J(n)]^\T \in \mathbb{C}^J$ and $\vn(n) \in \mathbb{C}^M$ denote the signal sent by each of the $J$ users and the noise, respectively, whereas $\mh = [\vh_1, ..., \vh_J]$ contains the individual channels of the users $\vh_j \in \mathbb{C}^M$.
We assume that the noise is Gaussian with $\vn(n) \sim \NC (\bm{0}, \mc_\vn = \sigma^2\I_M)$.
For the task of channel estimation, we consider a channel coherence interval larger than the number of snapshots $N$, i.e., the channel is constant over all snapshots. 

In conventional channel estimation schemes, some of the signals sent by each users consist of $N_p$ uplink pilots. The pilots sent by each user are known to the \ac{BS}. Hence, the received observations at the \ac{BS} side can be formulated as
\begin{align}
	\my=\left[\my_p,\my_d\right] = \mh\left[\matp, \md\right] +\mn= \mh\mx +\mn,
	\label{Eq:AllObservations}
\end{align}
where $\my_p\in\mathbb{C}^{M\times N_p}$, $\my_d\in\mathbb{C}^{M\times N-N_p}$, $\matp\in\mathbb{C}^{J\times N_p}$, and $\md\in\mathbb{C}^{J\times N-N_p}$ denote the received pilot observations, received data observations, sent pilots, and sent data symbols, respectively.
In order to fully illuminate the channels, the number of pilots is, at minimum, the number of users $N_p\geq J$, and orthogonal pilots are used. 
We set $N_p=J$, and utilize Discrete Fourier transform (DFT) pilot sequences.
After decorrelating the orthogonal pilot sequences the received pilot observations simplify to
\begin{align}
	\my_p = \mh +\mn. \label{eq:pilotsys}
\end{align}
Thus, the pilot observations of each user do not depend on the pilots sent by the remaining users.
This enables to consider channel estimation from a per user perspective in the subsequent discussions. 
For reasons of simpler readability, the index for the respective user is therefore no longer given in the following.
Consequently, we denote the pilot observation of a user as 
\begin{align}
	\vy_p = \vh + \vn, \label{eq:pilotsys_su}
\end{align}
with $\vn \sim \NC(0,\mc_\vn=\sigma^2\I_M)$. 

\subsection{Measurement Campaign}
\label{sec:meas}


We work with a training dataset $\mathcal{H} = \{\vh_t\}^T_{t=1}$ containing $T$ channel samples representing the entire BS cell's user channel distribution. 
Usually, simulation tools with sophisticated models are used to generate such datasets. 
However, these models capture real-world \ac{CSI} characteristics up to some extent.
To address this, we utilize real-world data from a measurement campaign conducted at the Nokia campus in Stuttgart, Germany, during October/November 2017, 
cf.~\cite{Turan2022}. 
The receive antenna with a uniform rectangular array (URA) comprised of $4$ vertical ($\lambda$ spacing) and $16$ horizontal ($\lambda/2$ spacing) single polarized patch antennas operating at a carrier frequency of $2.18$ GHz was mounted on a rooftop approximately $20$ meters above the ground. 
For further details, we refer the reader to~\cite{Turan2022}. 
%
%
With measurements performed at a high \ac{SNR} ranging from $20$ dB to $30$ dB, the measured channels are considered as representation of ground truth. However, for our investigation, we intentionally introduce artificial noise by corrupting the measured channels with additive white Gaussian noise (AWGN) at specific \acp{SNR} to obtain noisy observations.
It should be noted that our study focuses on a scenario where the coherence interval of the covariance matrix aligns with the channel's time scale, meaning the channel covariance matrix changes simultaneously with the channel.

%% file: method.tex
\section{Data-aided Gaussian Mixture Model}

\subsection{\ac{GMM}-based Channel Estimator}
\label{subsec:GMM}

Commonly, the pilot observation $\vy_p$ is only considered for channel estimation.
The \ac{MSE} optimal estimator is given by the \ac{CME} $\hat{\vh} = \mathbb{E}\left[\vh\mid\vy_p\right]$,
which generally can not be computed analytically. 
Also, the PDF of $\vh$ is usually not available.
Nevertheless, an estimator was introduced in \cite{Koller2022} which approximates the \ac{CME} utilizing a \ac{GMM}.
The PDF of $\vh$ is approximated by a \ac{GMM} as
\begin{align}
	f_\vh^{(K)}(\vh) = \sum_{k=1}^{K} p(k) \NC(\vh;\vmu_k,\mc_k), \label{eq:GMM_h}
\end{align}
where $p(k)$, $\vmu_k$ and $\mc_k$ are the mixing coefficients, means,
and covariances of the $k$-th \ac{GMM} component, respectively.
The fitting of the components is accomplished with the well-known expectation-maximization (EM) algorithm~\cite{Bishop2006} given a set $\mathcal{H} = \{\vh_t\}^T_{t=1}$ of $T$ channel
samples as training data.
Due to the Gaussianity of the noise with noise covariance matrix $\mc_\vn$ the PDF of the pilot observations $\vy_p$ can be approximated by
\begin{align}
	f_{\vy_p}^{(K)}(\vy_p) = \sum_{k=1}^{K} p(k) \NC(\vy_p;\vmu_k,\mc_k + \mc_\vn).
\end{align}
Leveraging these approximations of the \acp{PDF} of $\vh$ and $\vy_p$, a convex combination of \ac{LMMSE} estimates can be used to calculate a channel estimate as~\cite{Koller2022}
\begin{align}
	\hat{\vh}_\text{GMM} = \sum_{k=1}^K p(k\mid\vy_p)\hat{\vh}_{\text{GMM},k},
\end{align}
where $p(k\mid\vy)$ is the so-called responsibility of the $k$-th component, i.e., the probability that the $k$-th component is responsible for the observation $\vy_p$,~cf.~\cite{Koller2022}, and 
\begin{align}
	\hat{\vh}_{\text{GMM},k} = \mc_k \left(\mc_k + \mc_\vn\right)^{-1}\left(\vy_p - \vmu_k\right) + \vmu_k.
\end{align}
In this case, only the pilot observation of the current coherence interval can be utilized for channel estimation.

\subsection{Maximum Likelihood Subspace Estimation}

The received data symbols are also transmitted over the same channel and, hence, can be used to enhance the \ac{CSI} estimation quality.
Let us first consider the problem from a blind perspective, where no information about the sent symbols is available on the receiver side.
For this scenario and in view of \eqref{Eq:AllObservations}, the \ac{ML} estimator can be formulated as the solution of
\begin{align}
	 \min_{\mh,\mx} \sum_{n=1}^{N}\|\vy(n) - \mh \vx(n)\|_2^2, \label{eq:dmle}
\end{align}
which can be rewritten as~\cite{DeCarvalho1997a}
\begin{align}
	\max_\mh \;\tr\left(\matp_\mh\hat{\mc}_{\vy\mid\mh}\right), \label{eq:dmle2}
\end{align}
where $\matp_\mh= \mh (\mh^\He\mh)^{-1}\mh^\He$ and $\hat{\mc}_{\vy\mid\mh} = \frac{1}{N}\my\my^\He. \label{eq:smplcov}$
The maximization in \eqref{eq:dmle2} is solved 
by setting $\matp_\mh$ equal to $\mv \mv^\He$
with $\mv = [\vv_1,...,\vv_J]$ holding the $J$ dominant orthogonal eigenvectors of the receive sample covariance matrix $\hat{\mc}_{\vy\mid\mh}$.
Accordingly, we can see that $\text{range}(\mv)$ contains the estimated channels, cf.~\cite{Deng2020}.
One should note that infinitely many solutions exist, and the blind \ac{ML} estimator can only estimate the subspace containing the solutions.

\subsection{Subspace \ac{GMM} Channel Estimator}
\label{subsec:subGMM}

Using the information in $\text{range}(\mv)$, we can solve the estimation within the subspace.
For this, the pilot system model in \eqref{eq:pilotsys_su} is projected into the $J$-dimensional subspace with
\begin{align}
	\mv^\He\vy_{p} &= \mv^\He\vh + \mv^\He \vn = \vh^\prime + \vn^\prime. \label{eq:sub}
\end{align}
After solving the estimation in the subspace for $\vh^\prime$, the solution can be transformed back using
\begin{align}
	\hat{\vh}_\text{sub} = \mv \hat{\vh}^\prime. \label{eq:sub2}
\end{align}
Utilizing the covariance matrix $\mc_{\vn^\prime} = \mv^\He\mc_\vn\mv = \sigma^2\I_J$ based on \eqref{eq:sub}, we can employ the \ac{GMM} estimator from Section \ref{subsec:GMM} to solve the subspace estimation with the component-wise \ac{LMMSE} as
\begin{align}
	\hat{\vh}_{\text{sub. GMM},k} = \;&\mv\mv^\He\mc_k\mv \left(\mv^\He\mc_k\mv + \sigma^2\I_J\right)^{-1}\nonumber\\
	&\times\left(\mv^\He\vy_p - \mv^\He\vmu_k\right) + \mv\mv^\He\vmu_k, \label{eq:subGMM}
\end{align}
and the corresponding responsibilities
\begin{align}
    p(k\mid\vy_p) = \frac{p(k)\NC\left(\vy_p;\mv^\He\vmu_k,\mv^\He\mc_k\mv + \sigma^2\I_J\right)}{\sum_{i=1}^K p(i)\NC\left(\vy_p;\mv^\He\vmu_i,\mv^\He\mc_i\mv + \sigma^2\I_J\right)}.
\end{align}


\subsection{Projected \ac{GMM} Channel Estimator}
\label{subsec:projGMM}

An alternative approach uses the orthogonal subspace projection as a preprocessing filter
\begin{align}
	\tilde{\vy} = \matp_\mh\vy_p = \vh + \matp_\mh \vn = \vh + \tilde{\vn}.
\end{align}
The equality is because a perfect projection $\matp_\mh$ does not affect $\vh$.
To formulate the \ac{GMM} estimator after the projection, we need to calculate the statistic of the noise $\tilde{\vn}$ with
\begin{align}
	\mc_{\tilde{\vn}} = \;&\mathbb{E}_{\mh,\vn}\left[\tilde{\vn}\tilde{\vn}^\He\right] = \mathbb{E}_\mh\left[\sigma^2\matp_\mh\right] \label{eq:cov3}. 
\end{align}
To get an intuitive understanding of \eqref{eq:cov3}, let us consider a scenario involving spatially uncorrelated channels, meaning that path gains and channel directions are uncorrelated. 
This is the case when users are uniformly distributed over the directions, resulting in a channel covariance matrix that is a scaled identity~\cite[Def. 2.3]{Bjoernson2017}. 
This setup shares similarities with the widely used model of i.i.d. Rayleigh fading.
In such a case, the matrices with the eigenvectors of the sample covariance matrix of such channels are distributed with Haar measure~\cite[Chap. 1]{Milman1986}, i.e., uniformly distributed on the manifold of unitary matrices, which results in 
\begin{align}
	\mathbb{E}_\mh\left[\matp_\mh\right] = \frac{J}{N}\I_M.
\end{align}
In real-world scenarios, e.g., the measurement campaign detailed in Section \ref{sec:meas}, the assumption of spatial uncorrelated channels may no longer hold. Nevertheless, we can assume it for approximating the new noise covariance matrix $\mc_{\tilde{\vn}}$.
Hence, we approximate \eqref{eq:cov3} as
\begin{align}
	\mc_{\tilde{\vn}} \approx \sigma^2\frac{J}{M}\I_M.
\end{align}
In Section \ref{sec:sim}, we will assess the performance of this approximation.
We can now formulate the projected \ac{GMM} estimator as
\begin{align}
	\hat{\vh}_{\text{proj. GMM},k} = \;&\mc_k \left(\mc_k + \sigma^2\frac{J}{M}\I_M\right)^{-1}\left(\tilde{\vy} - \vmu_k\right) + \vmu_k \label{eq:projGMM}
\end{align}
with the associated responsibilities
\begin{align}
    p(k\mid\tilde{\vy}) = \frac{p(k)\NC\left(\tilde{\vy};\vmu_k,\mc_k + \sigma^2\frac{J}{M}\I_M\right)}{\sum_{i=1}^K p(i)\NC\left(\tilde{\vy};\vmu_i,\mc_i + \sigma^2\frac{J}{M}\I_M\right)}.
\end{align}

\subsection{Complexity Analysis}

The standalone \ac{GMM} estimator from \cite{Koller2022} precomputes the filters used for the individual components and, hence, only exhibits a complexity of $\mathcal{O}(KM^2)$. Also, the calculations of the $K$ components can be parallelized.
For all data-aided methods, the calculation of the subspace needs $\mathcal{O}(JM^2)$ since for the solution of \eqref{eq:dmle2} we are only interested in the eigenvectors of the $J$ largest eigenvalues.
When using the subspace \ac{GMM}, we have an additional complexity of $\mathcal{O}(K(M^2 + JM^2 + J^3))$ for the computation of the $K$ \ac{LMMSE} estimates in \eqref{eq:subGMM}. One should note that the $K$ components can be parallelized, but the filters of each of the $K$ components need to be computed for every $\mv$, including the inverse of a $J\times J$ matrix.
In contrast, for the projected version in \eqref{eq:projGMM} the complexity is only $\mathcal{O}(KM^2 + JM^2)$. 
Here, the $K$ components can be parallelized, and the filters can be precomputed.

%% file: estimators.tex
\section{Baseline Channel Estimators}




To compare our methods, the following baseline channel estimators are considered.
Based on the found subspace $\text{range}(\mv)$ we can formulate the \ac{ML} estimation problem of $\vh$ as
\begin{align}
	\min_{\bm{\alpha}} \|\vy_{p} - \mv\bm{\alpha} \|^2, \label{eq:pilotDML}
\end{align}
where $\bm{\alpha}$ denotes the coefficients such that $\hat{\vh}_{\text{ML}} = \mv\bm{\alpha}$.
The closed-form solution for the pilot-based \ac{ML} channel estimate 
is therefore
\begin{align}
	\hat{\vh}_{\text{ML}} = \mv \mv^\He \vy_{p}.
\end{align}
This can be interpreted as the subspace-adjusted version of the conventional least squares (LS) channel estimator given as $\hat{\vh}_\text{LS} = \vy_p$.


Another estimator is based on the sample covariance matrix, which we can compute from a training data set $\mathcal{H}$ with samples from the whole scenario according to $\mc = \frac{1}{|\mathcal{H}|} \sum_{\vh\in \mathcal{H}} \vh\vh^\He$,
which can be used to formulate the LMMSE estimator as
\begin{align}
	\hat{\vh}_{\text{s-cov}} = \mc\left(\mc + \sigma^2\I\right)^{-1}\vy_{p}.
\end{align}
Similar as outlined in Section \ref{subsec:subGMM} and Section \ref{subsec:projGMM}, we can adjust the sample covariance-based estimator using 
$\mv$ as
\begin{align}
	\hat{\vh}_{\text{sub. s-cov}} =\mv\mv^\He\mc\mv \left(\mv^\He\mc\mv + \sigma^2\I_J\right)^{-1}\mv^\He\vy_{p} ,
\end{align}
and
\begin{align}
	\hat{\vh}_{\text{proj. s-cov}} =\mc \left(\mc + \sigma^2\frac{J}{M}\I_M\right)^{-1}\matp_\mh\vy_{p}.
\end{align}

%% file: simulations.tex
\section{Numerical Simulations}
\label{sec:sim}

\begin{figure}[t]
	\centering
	\begin{tikzpicture}
		\centering
		\begin{semilogyaxis}[
			width=\columnwidth,
			height=5.2cm,
			ylabel={\footnotesize NMSE},
			xlabel={\footnotesize SNR [dB]},
			xmin=-15,
			xmax=20,
			ymin=0.001,
			ymax=2,
			grid=both,
			xtick={-20,-15,-10,-5,0,5,10,15,20,25,30,35,40},
			ytickten={-6,...,3},
			legend columns=4,
			legend style={at={(1.00,1.03)}, anchor=south east, font=\scriptsize},
			]		
			\addplot[mark=x,mark size=2pt,line width=1pt, color=black, dashed, mark options=solid] table [col sep=comma] {data/nmse_ura64_measurement_ls.csv};
			\addlegendentry{LS}		
			
			
			\addplot[mark=pentagon,mark size=1.8pt, line width=1pt, color=gray, dashed, mark options=solid] table [col sep=comma] {data/nmse_ura64_measurement_smpl_cov_quadriga.csv};
			\addlegendentry{s-cov}
			
			\addplot[mark=triangle,mark size=1.5pt, line width=1pt, color=green!50!gray, dashed, mark options={solid, rotate=180}] table [col sep=comma] {data/nmse_ura64_measurement_smpl_cov_heuristic_quadriga_200snaps_8ue.csv};
			\addlegendentry{proj. s-cov}
			\addplot[mark=diamond,mark size=1.8pt, line width=1pt, color=brown, dashed, mark options=solid] table [col sep=comma] {data/nmse_ura64_measurement_smpl_cov_subspace_quadriga_200snaps_8ue.csv};
			\addlegendentry{sub. s-cov}
			\addplot[mark=star,mark size=2pt, line width=1pt, color=red, dashed, mark options=solid] table [col sep=comma] {data/nmse_ura64_measurement_dmle_quadriga_8ue_200snaps.csv};
			\addlegendentry{ML}
			
			\addplot[mark=o,mark size=1.5pt, line width=1pt, color=orange, dashed, mark options=solid] table [col sep=comma] {data/nmse_ura64_measurement_gmm_quadriga.csv};
			\addlegendentry{GMM}	
			
			\addplot[mark=triangle,mark size=1.5pt, line width=1pt, color=blue] table [col sep=comma] {data/nmse_ura64_measurement_gmm_heuristic_quadriga_8ue_200snaps.csv};
			\addlegendentry{proj. GMM}
			\addplot[mark=square,mark size=1.5pt, line width=1pt, color=cyan] table [col sep=comma] {data/nmse_ura64_measurement_gmm_subspace_quadriga_8ue_200snaps.csv};
			\addlegendentry{sub. GMM}

		\end{semilogyaxis}
	\end{tikzpicture}
	\caption{
	NMSE over the \ac{SNR} for given channel estimations based on $N=200$ data observations and one pilot per user in a $J=8$ user scenario.
	}
	\label{fig:nmse_meas_SNR}
\end{figure}
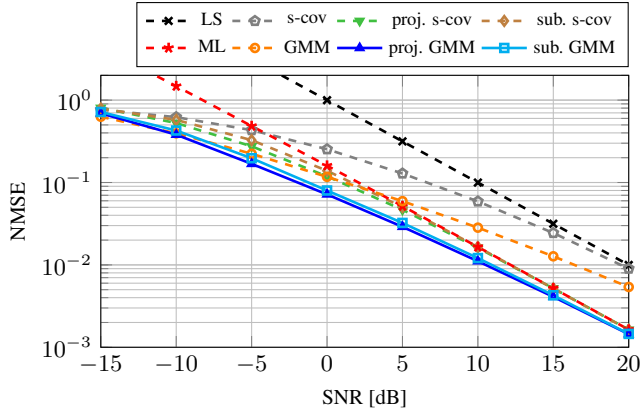

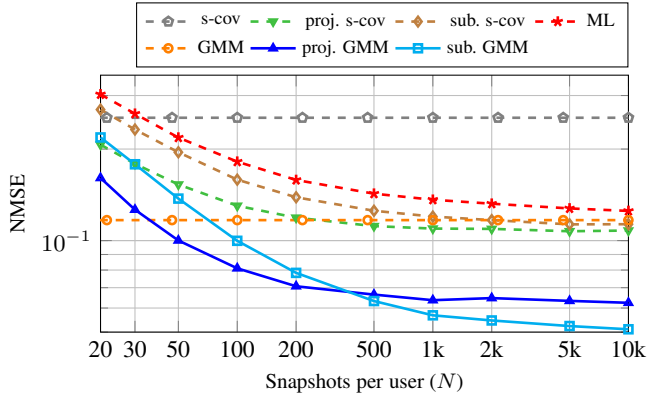
\begin{figure}[t]
	\centering
	\begin{tikzpicture}
		\centering
		\begin{loglogaxis}[
			width=\columnwidth,
			height=5cm,
			ylabel={\footnotesize NMSE},
			xlabel={\footnotesize Snapshots per user ($N$)},
			xmin=20,
			xmax=10000,
			ymin=0.05,
			ymax=0.35,
			grid=both,
			xtick={2,3,5,10,20,30,50,100,200,500,1000,2000,5000,10000},
			xticklabels={2,3,5,10,20,30,50,100,200,500,1k,2k,5k,10k},
			ytickten={-6,...,3},
			legend columns=4,
			legend style={at={(1.00,1.03)}, anchor=south east, font=\scriptsize},
			]		
			
			
			%
								
			\addplot[mark=pentagon, mark size=1.8pt, line width=1pt,domain=1:10000, color=gray, dashed, mark options=solid,mark repeat=2] {0.2535094009849667707};
			\addlegendentry{s-cov}
			
			\addplot[mark=triangle,mark size=1.5pt, line width=1pt, color=green!50!gray, dashed, mark options={solid, rotate=180}] table [col sep=comma] {data/nmse_ura64_measurement_smpl_cov_heuristic_quadriga_0dB_8ue_snaps.csv};
			\addlegendentry{proj. s-cov}
			
			\addplot[mark=diamond,mark size=1.8pt, line width=1pt, dashed, color=brown, mark options=solid] table [col sep=comma] {data/nmse_ura64_measurement_smpl_cov_subspace_quadriga_0dB_8ue_snaps.csv};
			\addlegendentry{sub. s-cov}
			
			\addplot[mark=star,mark size=2pt, line width=1pt, color=red, dashed, mark options=solid] table [col sep=comma] {data/nmse_ura64_measurement_dmle_quadriga_0dB_8ue_snaps.csv};
			\addlegendentry{ML}
			\addplot[mark=o, mark size=1.5pt,domain=1:10000, line width=1pt, color=orange, dashed, mark options=solid,mark repeat=2] {0.1167661725361357344};
			\addlegendentry{GMM}

			\addplot[mark=triangle,mark size=1.5pt, line width=1pt, color=blue, mark options=solid] table [col sep=comma] {data/nmse_ura64_measurement_gmm_heuristic_quadriga_0dB_8ue_snaps.csv};
			\addlegendentry{proj. GMM}
			
			\addplot[mark=square,mark size=1.5pt, line width=1pt, color=cyan, mark options=solid] table [col sep=comma] {data/nmse_ura64_measurement_gmm_subspace_quadriga_0dB_8ue_snaps.csv};
			\addlegendentry{sub. GMM}
	
		\end{loglogaxis}
	\end{tikzpicture}
	\caption{
	NMSE over the number of data observations for given channel estimations based on $N$ data observations and one pilot per user at \ac{SNR} $=0$ dB in a $J=8$ user scenario.
	}
	\label{fig:nmse_meas_snaps}
\end{figure}

We normalize the channel realizations with $\mathbb{E}\left[\|\vh\|^2\right]=N$ such that we can define the \ac{SNR} $=\frac{1}{\sigma^2}$.
Given $T$ channel estimates $\{\hat{\vh}_t\}^T_{t=1}$ of the test samples $\{\vh_t\}^T_{t=1}$, we can define the normalized MSE (NMSE) as $\frac{1}{NT}\sum_{t=1}^T \|\vh_t - \hat{\vh}_t\|^2.$
In our simulations, we use $T = 10^3$ channel samples stemming from the measurement campaign for evaluating the performances of the different channel estimators. In particular, we compare the two adapted \ac{GMM} estimators (``sub. GMM'' and ``proj. GMM'') with the \ac{GMM} estimator from~\cite{Koller2022} and the related estimators described above. We use $1.5\cdot10^5$ training samples from the measurement campaign to fit the \ac{GMM}. Each \ac{GMM} variant uses the same fitted \ac{GMM}, with $K=64$ components.
The ``s-cov'' variants (``sub. s-cov'' and ``proj. s-cov'') utilize the same training samples.
The number of \ac{BS} antenna is $M=64$ as explained in Section \ref{sec:meas}.
The chosen number of users is $J=8=M/8$, which is a representative operating point.
Also, if not stated otherwise, the number of snapshots is set to $N=200$, corresponding to a scenario that allows high mobility, e.g., up to $135$ kph, and high channel dispersion, c.f.~\cite[Chap. 2.1]{Bjoernson2017}.
For the sent symbols we choose Gaussian symbols with $x_j(n) \sim \NC (0, P_j=1/J)$ such that $\sum_{j=1}^J P_j =1$. 
The sent symbols in real systems stem from a discrete constellation, e.g., QPSK.
For our work, we stick to Gaussian symbols since a discrete symbol constellation does not affect the qualitative results of the simulations.

Fig. \ref{fig:nmse_meas_SNR} illustrates the performance of different channel estimation methods with respect to the \ac{SNR}. 
The projected \ac{GMM} performs the best for most \ac{SNR} values, closely followed by the subspace \ac{GMM}.
Furthermore, as the \ac{SNR} decreases, the \ac{GMM} variants approach the performance of the standard \ac{GMM}, while the sample covariance variants similarly converge towards the basic sample covariance estimator.
For high \ac{SNR} values, all data-aided variants approximate the ML estimator.
Interestingly, the \ac{GMM} variants outperform the ML estimator.
A notable observation is that, in the mid-\ac{SNR} range, the data-aided \ac{GMM} variants outperform all related estimators by roughly $3$ dB.

The accuracy of the estimated subspace influences the performance of data-aided variants. 
Figure \ref{fig:nmse_meas_snaps} shows that for a low number of snapshots, i.e., less accurate estimation of the subspace, the projected \ac{GMM} variant performs the best and is only surpassed by the standard \ac{GMM} approach when the number of snapshots is $30$ or fewer.
However, as the number of observed data snapshots increases to $500$ or more, the subspace \ac{GMM} outperforms all other estimators and appears to converge to a lower error level for large numbers of snapshots.
Interestingly, the data-aided sample covariance-based methods do not perform significantly better than the standard \ac{GMM} estimator, irrespective of the number of snapshots.

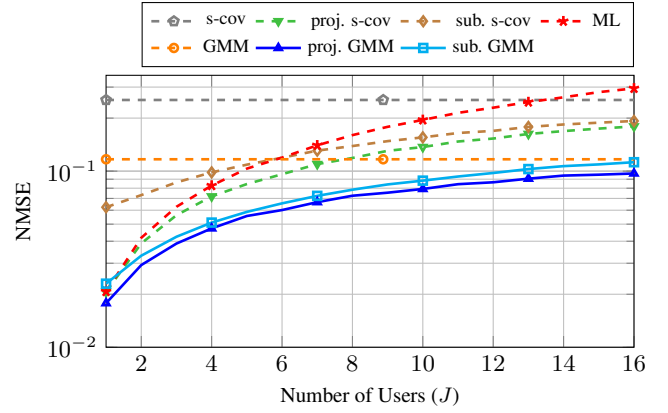
\begin{figure}[t]
	\centering
	\begin{tikzpicture}
		\centering
		\begin{semilogyaxis}[
			width=\columnwidth,
			height=5.2cm,
			ylabel={\footnotesize NMSE},
			xlabel={\footnotesize Number of Users ($J$)},
			xmin=1,
			xmax=16,
			ymin=0.01,
			ymax=0.35,
			grid=both,
			ytickten={-6,...,3},
			legend columns=4,
			legend style={at={(1.00,1.03)}, anchor=south east, font=\scriptsize},
			]		
			

			\addplot[mark=pentagon,mark size=1.8pt,line width=1pt,domain=1:64, color=gray, dashed, mark options=solid,mark repeat=3] {0.2535094009849667707};
			\addlegendentry{s-cov}
			
			\addplot[mark=triangle,mark size=1.5pt, line width=1pt, color=green!50!gray, dashed, mark options={solid, rotate=180},mark repeat=3] table [col sep=comma] {data/nmse_ura64_measurement_smpl_cov_heuristic_0dB_quadriga_200.csv};
			\addlegendentry{proj. s-cov}
			\addplot[mark=diamond,mark size=1.8pt, line width=1pt, color=brown, dashed,mark options=solid,mark repeat=3] table [col sep=comma] {data/nmse_ura64_measurement_smpl_cov_subspace_0dB_quadriga_200.csv};
			\addlegendentry{sub. s-cov}
			
			\addplot[mark=star,mark size=2pt, line width=1pt, color=red, dashed, mark options=solid, mark repeat=3] table [col sep=comma] {data/nmse_ura64_measurement_dmle_0dB_quadriga_200snaps.csv};
			\addlegendentry{ML}

			\addplot[mark=o,mark size=1.5pt,domain=1:64, line width=1pt, color=orange, dashed, mark options=solid, mark repeat=3] {0.1167661725361357344};
			\addlegendentry{GMM}

			\addplot[mark=triangle,mark size=1.5pt, line width=1pt, color=blue, mark options=solid, mark repeat=3] table [col sep=comma] {data/nmse_ura64_measurement_gmm_heuristic_0dB_quadriga_200snaps.csv};
			\addlegendentry{proj. GMM}
			
			\addplot[mark=square,mark size=1.5pt, line width=1pt, color=cyan, mark options=solid,mark repeat=3] table [col sep=comma] {data/nmse_ura64_measurement_gmm_subspace_0dB_quadriga_200snaps.csv};
			\addlegendentry{sub. GMM}

		\end{semilogyaxis}
	\end{tikzpicture}
	\caption{
	NMSE over the number of users for given channel estimations based on $N=200$ data observations and one pilot per user at \ac{SNR} $=0$ dB.
	}
	\label{fig:nmse_meas_user}
\end{figure}

The subspace dimension is directly influenced by the number of users in the system.
Fig. \ref{fig:nmse_meas_user} shows the performances for different numbers of users.
As the number of users approaches the number of \ac{BS} antennas, all estimators converge to their respective standalone pilot versions since the subspace projection results in the identity.
Moreover, in the case of a single user, all data-aided variants, except for the subspace sample covariance estimator, exhibit similar performance. 
However, with an increasing number of users, the differences among the data-aided variants become more pronounced.
The preferred operating regime in multi-user MIMO is $J \leq M/4=16$~\cite[Chap. 1.3.3]{Bjoernson2017}. 
As can be seen in Fig. \ref{fig:nmse_meas_user}, especially in this regime ($J \leq 16$), the benefit of using the proposed data-aided versions results in superior channel estimates. 



%% file: conclusion.tex
\vspace{3.5pt}
\section{Conclusion}


This work showed how the received data symbols can be utilized at the \ac{BS} to enhance the channel estimation quality in a multi-user scenario.
We introduced two different approaches based on the GMM channel estimation framework. 
Both methods exploit the estimated subspace derived from the dominant eigenvectors of sample covariance matrices constructed using the received symbols.
Extensive simulations based on real-world measurement data demonstrated the superior estimation performance of both proposed methods. 
In future work, we aim to extend the findings to a multi-cell scenario where pilot contamination plays a critical role~\cite{Jose2009}. Additionally, we intend to investigate the utilization of imperfect training data~\cite{Fesl2023} and the impact of structured covariance matrices~\cite{Turan2022} on the performance of our proposed data-aided \ac{GMM}-based estimators.